\documentclass[aps,prd,nofootinbib,superscriptaddress,preprint,showkeys,showpacs,preprintnumbers]{revtex4}
\usepackage{color}
\usepackage{slashed}
\usepackage{graphics}
\usepackage{graphicx}
\usepackage{dcolumn}
\usepackage{subfigure}
\usepackage{mathrsfs}
\usepackage{bm}
\usepackage{amsmath,amssymb,epsfig}
\usepackage{float}
\allowdisplaybreaks[3]

\newcommand{\bea}{\begin{eqnarray*}}
\newcommand{\eea}{\end{eqnarray*}}
\newcommand{\bean}{\begin{eqnarray}}
\newcommand{\eean}{\end{eqnarray}}

\newcommand{\hsp}{\hspace{0.1mm}}

\newcommand{\pa}{\partial}

\begin{document}

\title{Self-consistent Effective-one-body theory for spinless  binaries   based on  post-Minkowskian approximation I: Hamiltonian and decoupled equation for  $\psi^B_{4}$}

\author{Jiliang {Jing}\footnote{ jljing@hunnu.edu.cn}}
 \affiliation{Department of Physics, Key Laboratory of Low Dimensional Quantum Structures and Quantum Control of Ministry of Education, and Synergetic Innovation
Center for Quantum Effects and Applications, Hunan Normal
University, Changsha, Hunan 410081, P. R. China}
\affiliation{Center for Gravitation and Cosmology, College of Physical Science and Technology, Yangzhou University, Yangzhou 225009, P. R. China}

\author{Shuai Chen}
 \affiliation{Department
of Physics, Key Laboratory of Low Dimensional Quantum Structures and
Quantum Control of Ministry of Education, and Synergetic Innovation
Center for Quantum Effects and Applications, Hunan Normal
University, Changsha, Hunan 410081, P. R. China}

\author{Manman Sun}
\affiliation{Department
of Physics, Key Laboratory of Low Dimensional Quantum Structures and
Quantum Control of Ministry of Education, and Synergetic Innovation
Center for Quantum Effects and Applications, Hunan Normal
University, Changsha, Hunan 410081, P. R. China}

\author{Xiaokai He}
 \affiliation{School of Mathematics and
Statistics, Hunan First Normal University, Changsha
410205, China}

\author{Mengjie Wang} 
\affiliation{Department
of Physics, Key Laboratory of Low Dimensional Quantum Structures and
Quantum Control of Ministry of Education, and Synergetic Innovation
Center for Quantum Effects and Applications, Hunan Normal
University, Changsha, Hunan 410081, P. R. China}

\author{Jieci Wang\footnote{ jcwang@hunnu.edu.cn}} 
\affiliation{Department
of Physics, Key Laboratory of Low Dimensional Quantum Structures and
Quantum Control of Ministry of Education, and Synergetic Innovation
Center for Quantum Effects and Applications, Hunan Normal
University, Changsha, Hunan 410081, P. R. China}


\begin{abstract}
To build a self-consistent effective-one-body (EOB) theory, in which the  Hamiltonian,  radiation-reaction force  and waveform for the ``plus" and ``cross" modes of the gravitational wave should be based on the same  effective background spacetime,  the key step is to look for the decoupled equation for  $\psi^B_{4}=\ddot{h}_{+}-i\ddot{h}_{\times}$, which seems a very difficult task because there are non-vanishing tetrad components of the tracefree Ricci tensor for such spacetime. Fortunately,  based on an effective spacetime obtained in this paper by using the post-Minkowskian (PM) approximation,  we find the decoupled equation for $\psi^B_{4}$ by dividing the perturbation part of the metric into  the odd and even parities.  With the effective metric and decoupled equation at hand, we set up a frame of self-consistent EOB model for spinless  binaries. 

\end{abstract}


\maketitle

 \section{Introduction}

Gravitational radiation has attracted much attention since 1918 \cite{einstein18,BonVanMet62,Sac62, deAdeBoc00,KomSmiDun11,PerAldDel98,Jing2021,Jing2019}.
Quite a lot of gravitational wave (GW) events, which are generated by coalescing compact object binary systems, have been confirmed since the first direct detection of GW by Laser Interferometer Gravitational-Wave Observatory \cite{Abbott2016, Abbott2016(2),Abbott2017,Abbott2017(2),Abbott2017(3),Abbott2019,Abbott20211,Abbott20212,Abbott20213,Nitz2021}. In the detection of the GW events generated by coalescing binary systems, the gravitational waveform template plays a central role. It is well known that the basis of the gravitational waveform template is the theoretical models of gravitational radiation, and the key point to construct the theoretical models is to study the late dynamical evolution of a coalescing binary system of compact objects.

To investigate the general gravitational radiation generated by coalescing compact object binary systems, in 1999, Buonanno and Damour \cite{Damour1999} introduced, based on the post-Newtonian (PN) approximation, a novel approach to map the two-body problem onto an EOB problem, i.e. the motion of a test particle in an effective external spacetime. 
Based on the EOB theory with the PN approximation, Damour et al. provided an estimate of the gravitational waveform emitted throughout the inspiral, plunge and coalescence phases~\cite{Damour2000}, and determined the last stable circular orbit for general relativistic binaries~\cite{Damour2000(2)}. These studies were then generalized to the case of spinning black holes~\cite{Damour2001, Damour2006}. The EOB formalism is a successful theoretical model to describe the gravitational radiation emitted by binary black holes. Later on, the EOB inspiral-merger-ringdown waveforms were improved by calibrating the model to progressively more accurate numerical relativity simulations, spanning larger regions of the parameter space~\cite{Cook,Pan,Pan1,Damour2008,Damour20082, Boyle,Damour20083,Pan2,Pan3,Damour2009,Pan4,Bar,Pan5,Pan2014,Bohe,Damour2015,Pan1,Cao2017}, and played a vital role in the analysis of the gravitational wave signals in the LIGO/Virgo collaboration.

In 2016, Damour~\cite{Damour2016} presented another approach, by combining EOB theory with post-Minkowskian (PM), instead of PN, approximation. This implies, in the new model, the assumption that $v/c$  is a small quantity in the PN has been removed.

The Hamilton equations \cite{Damour2000}  for an EOB system can be  expressed as \begin{eqnarray}
&&\frac{dR}{d t} - \frac{\pa H[g_{\mu\nu}^{\text{eff}}]}{\pa P_R}
 = 0, \nonumber \\ &&  \frac{d \varphi}{d t} - \frac{\pa H[g_{\mu\nu}^{\text{eff}}]}{\pa P_\varphi}
 = 0, \nonumber \\
&& \frac{d P_R}{d t} + \frac{\pa H[g_{\mu\nu}^{\text{eff}}]}{\pa R}
= {\cal F}_R[g_{\mu\nu}^{\text{eff}}], \nonumber \\ && \frac{d P_\varphi}{d t} = {\cal F}_\varphi[g_{\mu\nu}^{\text{eff}}], \label{HEq}
\end{eqnarray}
where we use polar coordinates ($R, \varphi$) and the corresponding  conjugate momenta ($P_R, P_{\varphi} $). The rate of energy-loss along any orbits, in polar coordinates, is given by $\frac{d  E[g_{\mu\nu}^{\text{eff}}]}{d t}= \dot{R}\,{\cal F}_{R} [g_{\mu\nu}^{\text{eff}}]+
\dot{\varphi}\,{\cal F}_{\varphi}[g_{\mu\nu}^{\text{eff}}]\,.$ 
For quasi-circular motions, an excellent approximation to the radiation-reaction force is obtained by replacing the radial component simply by zero     \cite{Damour2000}. Then,  the radiation-reaction  force becomes 
\begin{eqnarray}\label{dEt}
{\cal F}_{\varphi}^{\rm circ}[g_{\mu\nu}^{\text{eff}}] \simeq \frac{1}{\dot{\varphi}}\frac{d  E[g_{\mu\nu}^{\text{eff}}]}{d t},
\end{eqnarray}
which indicates that, to get the radiation-reaction force for the ``plus" and ``cross" modes of the gravitational wave, we shall find the energy-loss rate for these modes.  

From the above discussions we know that, for a  self-consistent EOB theory, the 
Hamiltonian $H[g_{\mu\nu}^{\text{eff}}]$, the radiation-reaction force  $ {\cal F}_\varphi [g_{\mu\nu}^{\text{eff}}]$  and the waveform for the ``plus" and ``cross" modes should be based on the same effective background spacetime, $g_{\mu\nu}^{\rm eff}$. 

To build such a self-consistent EOB theory,  the key step is to look for the decoupled equation for the tetrad component of the perturbed Weyl tensor  $\psi^B_{4}=\ddot{h}_{+}-i\ddot{h}_{\times}$ in the effective spacetime, which seems impossible because there are non-vanishing tetrad components of the tracefree Ricci tensor  ($\phi_{00},$ $ \phi_{22} $ and $\phi_{11}$)  for such spacetime.  As we know,  before our study,  the decoupled equation for   $\psi^B_{4}$ was only obtained in the Schwarzschild and Kerr spacetimes because all tetrad components of the tracefree Ricci tensor are vanishing in the two kinds of spacetimes. Therefore, in previous studies on waveform template  based on the EOB theory, the Hamiltonian $H$ was based on the effective metric, but in some works used the radiation reaction force  and the waveform for the ``plus" and ``cross" modes based on the Schwarzschild or Kerr metric, and in some works used the Newtonian multipolar waveform which corresponds to post-Newtonian approximation metric but not to the effective metric. 

In this manuscript we will build a self-consistent EOB theory based on the PM approximation because we find, for an effective spacetime obtained in this paper by using the PM approximation, the decoupled equation for $\psi^B_{4}$ by dividing the perturbation part of the metric into  the odd and even parities. 

The rest of the paper is organized as follows. We first study the  energy map, effective metric and Hamiltonian for the EOB theory by using the scattering angles for both the real two-body  and the EOB systems in Sec. II. Then, in Sec. III, we look for the decoupled equation for the tetrad component of the perturbed Weyl tensor, and show how to find the rate of energy-loss,  radiation-reaction force and waveform based on $\psi^B_{4}$. Finally, we present our conclusions and discussions.

In this paper,  we take part of the geometric units with only $c=1$, which is suitable for the calculations in the PM framework.


\section{The  energy map, effective metric and Hamiltonian for the EOB theory }

To find the Hamiltonian for the EOB theory, we shall first obtain the  effective  metric and the energy map between the relativistic energy $\mathcal{E}$ of real two-body system and the relativistic energy $\mathcal{E}_0$ of EOB system. 
Explicit calculations have shown that, different from the action variable and the precession angle, the scattering angle contains all order terms except  $G^0$  order one within the framework of PM approximation.  This implies that the scattering angle takes more information than radial action variable and precession  angle. Therefore, we will set up the EOB theory by using the scattering angle.

\subsection{Scattering angle for the real two-body system}

As shown by Bern et al. \cite{Bern2019,Bern20192},  the conservative Hamiltonian of massive spinless binary system with the masses $m_1$ and $m_2$ is described by 
$
H(\vec{p},\vec{r})=\sqrt{|\vec{p}|^{\;2}+m_1^2}
+\sqrt{|\vec{p}|^{\;2}+m_2^2}+V(\vec{p},\vec{r}),
$
where 
$
V(\vec{p},\vec{r})=\sum_{i=1}^{\infty}c_i
\big{(}\frac{G}{|\vec{r}|}\big{)}^i,
$
 $\vec{r}$ is the distance vector between particles, $\vec{p}$ is the momenta, $i$ labels PM orders. The explicit expressions of $c_i\, (i=1,2,3)$ can be found in Refs.  \cite{Bern2019,Bern20192}.

For the real two-body system,  the scattering angle (3PM order can be found in Refs. \cite{Kalin,Kalin1} and 4PM order in Ref. \cite{Kalin2})   can be expressed as 
\begin{eqnarray}
\chi^{\text{rel}}=\chi^{\text{rel}}_1\frac{G}{J}+\chi^{\text{rel}}_2\Big(\frac{G}{J}\Big)^2+\chi^{\text{rel}}_3 \Big(\frac{G}{J}\Big)^3+\cdot \cdot\cdot  \cdot  \cdot \ \cdot  \;,
\end{eqnarray}
with
\begin{eqnarray}
&&\chi^{\text{rel}}_1 = 2 \,m_1\, m_2 \,\frac{2\gamma^2-1}{\sqrt{\gamma^2-1}}, \  \nonumber \\  &&\chi^{\text{rel}}_2=\frac{3\,\pi \,m_1^2\, m_2^2}{4}\, \frac{5\gamma^2-1}{\Gamma},\ \nonumber \\ 
&&\chi^{\text{rel}}_3= 2\, m_1^3\, m_2^3\,\sqrt{\gamma^2-1}\, P_{30} +\frac{2}{\pi}\chi^{\text{rel}}_1 \, \chi^{\text{rel}}_2-\frac{(\chi^{\text{rel}}_1)^3}{12} , \nonumber \\ 
&& \cdot \cdot \cdot \cdot \cdot\,  \cdot ,\nonumber  \label{2pmchi}
\end{eqnarray}
where  $\gamma = \frac{1}{2} \frac{\mathcal{E}^2-m_1^2-m_2^2}{m_1m_2} $,  $\Gamma =  \frac{\mathcal{E}}{m_1+m_2}$, $\nu=\frac{m_1 \, m_2}{(m_1+m_2)^2}$, $\mathcal{E}$ is the energy  of real two-body system,  and $
P_{30} = \frac{18\gamma^2-1}{2\,\Gamma^2}+\frac{8\,\nu\, (3+12\gamma^2-4\gamma^4)}{\Gamma^2\, \sqrt{|\gamma^2-1|}} \mbox{arcsinh}\sqrt{\frac{|\gamma-1|}{2}}
+\frac{\nu}{\Gamma^2}\big(1-\frac{103}{3}\gamma-18 \gamma^2-\frac{2}{3} \gamma^3+\frac{3 \,\Gamma\,(1-2\gamma^2)(1-5 \gamma^2)}{(1+\Gamma)(1+\gamma)}\big).
$

\subsection{Scattering angles  for  the EOB system  in the Schwarzschild-like coordinates}
We now calculate the effective scattering angle $\chi^{\text{eff}}$ in the Schwarzschild-like coordinates, which is deduced from the dynamics of a test particle with rest mass $m_0$ moving in an effective curved background spacetime describing by the metric $g_{\mu\nu}^{\rm eff}$ with the mass $M_0$. Following the spirit of EOB approach \cite{Damour1999,Damour2016}, the effective metric of a spinless real two-body system, in the PM approximation,  can be written as
\begin{eqnarray}
ds_{\rm eff}^2=A_0 dt^2-\frac{D_0^2}{A_0}dR^2- R^2(d\theta^2+\sin^2\theta d\phi^2),\label{Mmetric}
\end{eqnarray}
with 
\begin{eqnarray}
&& A_0=1-   2 GM_0/R+\sum_{i=2}^\infty a_i\big(GM_0/R\big)^i, \nonumber \\ && D_0=1+\sum_{i=2}^\infty d_i\big(GM_0/R\big)^i,
\end{eqnarray}
where we take $a_1=-2$ and $d_1=0$ by requiring that the effective metric at 1PM order should coincides with the Schwarzschild metric.

In the effective spacetime, the  scattering angle can be defined as
$
\chi^{\text{eff}}=-\pi-2  \int_{R_{\rm min}}^{\infty}\partial\sqrt{ \mathcal{R}_0}/\partial J_0 d R, 
$
where $\mathcal{R}_0=
D_0^2/A_0^2\mathcal{E}_0^2-D_0/A_0
\big(m_0^2+J_0^2/R^2\big)$ and  $R_\text{min}$  is the minimum distance  determined by setting vanishing of $\mathcal{R}_0$. Then by expanding $\chi^{\text{eff}}$ in terms of $G$, one obtains
\begin{eqnarray}\label{x} \chi^{\text{eff}}=\chi^{\text{eff}}_1\frac{G}{J_0}+\chi^{\text{eff}}_2 \Big(\frac{G}{J_0}\Big)^2+\chi^{\text{eff}}_3 \Big(\frac{G}{J_0}\Big)^3+\cdot \cdot \cdot  \cdot  \cdot \  \cdot ,
 \end{eqnarray}
with
\begin{align}
&\chi^{\text{eff}}_1=\frac{2 M_0 \left[2 \mathcal{E}_0^2-m_0^2 \right]}{\sqrt{\mathcal{E}_0^2-m_0^2}},  \nonumber \\
&\chi^{\text{eff}}_2=\frac{M_0^2 \pi}{4}\bigg{[}\mathcal{E}_0^2(15 - 3 a_2 + 2 d_2)+m_0^2( a_2-3 - 2 d_2)\bigg{]}, \nonumber  \\
&\chi^{\text{eff}}_3=\frac{2\chi^{\text{\text{\text{eff}}}}_1 \chi^{\text{eff}}_2}{\pi}-\frac{(\chi^{\text{eff}}_1)^3}{12} -\frac{M_0^3 \sqrt{\mathcal{E}_0^2-m_0^2} }{3}
[(3 - 3 a_2 - 2 a_3 \nonumber \\ &   - 2 d_2 + 4 d_3)     m_0^2 + 2 \mathcal{E}_0^2( 15 a_2 + 2(2 a_3 -  d_2 -  d_3) -27)],  \nonumber\\
&\cdot \cdot \cdot  \cdot  \cdot  \  \cdot . \nonumber \end{align}

\subsection{The  energy map, effective metric and Hamiltonian for the EOB theory}

Following Buonanno and Damour \cite{Damour1999,Damour2016}, the relations for the masses and angular momentums between EOB system and real two body system can be taken as
$
m_0=\frac{m_1m_2}{m_1+m_2},\ 
M_0=m_1+m_2, \ 
J_0=J.\label{i2}
$
By identifying the scattering angles for the EOB and real two body systems at the 1PM order,  we obtain the energy map 
\begin{eqnarray}\label{EMap}
\mathcal{E}_0&=\frac{\mathcal{E}^2-m_1^2-m_2^2}{2(m_1+m_2)}\label{energymap1}\;. \end{eqnarray}

The parameters ($a_i, \ d_i$) appearing in the effective metric can be found by identifying the scattering angles for the two systems order by order.  
However, there are two unknowns $a_i$ and $d_i$ for one  equation at each order $i$. Thus, one of them, say $d_i$, can be considered as undetermined parameter which will be fixed in section III. Then, the parameters $a_i$ are given by 
\begin{eqnarray} \label{parameters}
a_2&=&\frac{3(1- \, \Gamma)(1-5 \, \gamma^2)}{\Gamma\, (3\, \gamma^2-1 )}-d_2\, \frac{2\,  (1-  \gamma^2)}{ 3\, \gamma^2-1 }\, ,\nonumber \\ 
a_3&=&\frac{3}{2 }\Big[\frac{3- 2\, \Gamma-3 (15 -8\, \Gamma ) \gamma^2+6 ( 25-16 \, \Gamma ) \gamma^4}{\Gamma\, (4\, \gamma^2-1 ) (3\, \gamma^2-1 )}\nonumber \\ &-&\frac{2 P_{30}}{(4 \gamma^2-1 )}\Big]-d_2 \frac{4-34 \gamma^2+24 \gamma^4}{1-7  \gamma^2+12 \gamma^4 }+d_3\frac{ 2 (1-  \gamma^2)}{1-4 \gamma^2}, \nonumber \\    &&\cdot \cdot \cdot  \cdot  \cdot  \  \cdot .
     \end{eqnarray}

Then, by means of the effective metric, we can construct the effective Hamiltonian  using Buonanno-Damour's proposals  \cite{Damour1999,Damour2000}, which is given by 
\begin{eqnarray} \label{Heff}
H_{\rm eff}=m_0\sqrt{A_0\Big(1+\frac{A_0P_R^2}{m_0^2 D_0^2}+\frac{P_\varphi^2}{ m_0^2 R^2}+Q \Big)},  
\end{eqnarray}
where  $Q$ represents a post-geodesic (Finsler-type) contribution which, at least, is quartic in momenta \cite{Da1}. While by using Eq. (\ref{EMap}), the EOB Hamiltonian appearing in Eq. (\ref{HEq}) becomes 
\begin{eqnarray} \label{Heob}
H[g_{\mu\nu}^{\text{eff}}]=M_0 \sqrt{1+2\nu \left(\frac{H_{\rm eff}}{m_0} -1 \right)}\,.
\end{eqnarray}

\vspace{0.3cm}
\section{The rate of energy-loss,  radiation-reaction force and waveform}

To get the radiation-reaction force which can be derived from the energy-loss rate for the ``plus" and ``cross" modes of the gravitational wave,  we should find the decoupled equation for $\psi^B_{4}=\ddot{h}_{+}-i\ddot{h}_{\times}$.  In addition,  from  $\psi_4^B$ we can also read out the waveform  for the ``plus" and ``cross" modes of the gravitational wave.

\subsection{Decoupled gravitational wave equations}

A straightforward and natural way to obtain gravitational perturbation equations for the background spacetime described by the effective metric  $ g_{\mu \nu}^{\text{eff}}=g_{\mu \nu}^{A}$  is to start with a perturbation metric 
\begin{align}\label{Pmetric}
  g_{\mu \nu}=g_{\mu \nu}^{A}+\varepsilon h_{\mu \nu}^{B}\;,
\end{align}
where  $\varepsilon$ is a small quantity, the superscripts $A$ and $B$ denote the background and perturbation quantities, respectively.  To do perturbation theory using the tetrad formalism, we specify the perturbed Newman-Penrose tetrad by $l_\mu=l_\mu^A+\varepsilon l_\mu^B$, $n_\mu=n_\mu^A+\varepsilon n_\mu^B$ and $m_\mu=m_\mu^A+\varepsilon m_\mu^B$. All the tetrad components of the Weyl tensor, spin coefficients and intrinsic derivatives can then be written as  $\Psi_i=\Psi_i^A+\varepsilon \Psi_i^B$ ($i=0,1,2,3,4$),   $\phi_{ij}=\phi_{ij}^A+\varepsilon \phi_{ij}^B$ ($i=0,1,2$), and so on. Then, the complete set of perturbation equations is obtained from the Newman-Penrose equations  \cite{Newman} by keeping $\varepsilon$ up to the first order.

For the effective spacetime (\ref{Mmetric}),  the non-vanishing tetrad components of the tracefree Ricci tensor are   $\phi_{00},$ $ 
 \phi_{22} $ and $\phi_{11}$. To get the decoupled formula for perturbed Weyl tensor component $\Psi_4^B$,  we take
$ D_0=1.$
Then, $\phi_{00}=\phi_{22}=0 $ and the parameters ($a_i,\ d_i$) in Eq. (\ref{parameters}) are fixed.  Now, all non-vanishing spin coefficients, Weyl tensor component, and component of the tracefree  Ricci tensor are
\begin{eqnarray}
&& \rho^{A}=1/R, \ \ \ \mu^{A}=-A_0/(2R),\ \ \ \gamma^{A}=A_0'/4, \nonumber \\  && \alpha^{A}=-\beta^A=-cot\theta/(2\sqrt{2}R),\nonumber \\   &&\psi_{2}^{A}=(1/12R^{2})\big[2(A_0-1)-2 R A_0'+R^{2}A_0''\big], \nonumber \\  
 && \phi_{11}^{A}=(1/4 R^{2})(1-A_0+R^{2}A_0''/2).
\end{eqnarray}
Then,  the equation for $\psi_{4}^{B}$ is given by   
\begin{align}
&[(\Delta+3\gamma-\bar{\gamma}+4\mu+\bar{\mu})(D-\rho) -(\bar{\delta}+3\alpha+\bar{\beta}) (\delta+4\beta)\nonumber\\ 
&-3\psi_{2}+2\phi_{11} ]\psi^{B}_{4} =\frac{\hat{\kappa}}{2}T_4^B -4[(\Delta+2\mu+2\bar{\mu}+3\gamma-\bar{\gamma})\nonumber \\ & \times (\lambda^{B}\phi_{11})+\nu^{B}\bar{\delta}\phi_{11}],  \label{Psi4}
\end{align}
with
\begin{align}
T_4^B&=(\Delta+3\gamma-\bar{\gamma}+4\mu+\bar{\mu})[(\bar{\delta}+2\alpha)T^B_{n\bar{m}}  -(\Delta+2\gamma\nonumber \\ &-2\bar{\gamma}+\bar{\mu})T^B_{\bar{m}\bar{m}}]+(\bar{\delta}+3\alpha+\bar{\beta})[(\Delta+2\gamma+2\bar{\mu})T^B_{n\bar{m}}\nonumber \\ &-(\bar{\delta}+2\bar{\beta}+2\alpha)T^B_{nn}],
\end{align}
where all quantities without superscripts denote the background quantities,  
$T^B_{n\bar{m}}=\bar{\phi}_{12}^B/4(\pi G)$, $T^B_{\bar{m}\bar{m}}=\bar{\phi}_{02}^B/(4\pi G)$, and $T^B_{nn}=\phi_{22}^B/(4\pi G)$. Note that the above equation still includes $\lambda^{B}$ and $\nu^{B}$, which implies that $\psi_{4}^B$ still couples with other unknown quantities.

To get the decoupled equation for $\psi_{4}^B$, we rewrite 
Eq. (\ref{Pmetric}) as 
\begin{eqnarray}g_{\mu\nu}=g_{\mu\nu}^{A}+\varepsilon h_{\mu\nu}^{BE}+\varepsilon h_{\mu\nu}^{BO},\end{eqnarray}
i.e.,  we divide the metric perturbation into  the odd and even parities \cite{Thompson}
\begin{eqnarray}
h_{\mu\nu}^{BE}=\begin{pmatrix}
A \, S&-D\,  S &-R \, B  \frac{\partial S}{\partial{\theta}} &-R \, B \, \frac{\partial S}{\partial{\varphi}}\\
\text{Sym}\ \ &K \,  S &R \, H \, \frac{\partial S }{\partial{\theta}}&R\,  H \, \frac{\partial S}{\partial{\varphi}}\\
\text{Sym}\ \ &\text{Sym}\ \  &R^{2} \, E_{F}^{+} &R^{2}\, F \,  P_{S1}\\
\text{Sym}\ \  &\text{Sym}\ \  & \text{Sym}\ \  &R^{2}\,  sin\theta^{2}\,  E_{F}^{-}
\end{pmatrix},\nonumber \\
h_{\mu\nu}^{BO}=\begin{pmatrix}
0 &0 &\frac{R C}{sin\theta}\frac{\partial S}{\partial{\varphi}} &-R C sin\theta \frac{\partial S}{\partial {\theta}}\\
0 &0 &-\frac{R J}{sin\theta} \frac{ \partial S}{\partial {\varphi}} &R  J sin\theta \frac{\partial S}{\partial{\theta}}\\
\text{Sym}    &\text{Sym}\   & -\frac{R^{2}  G}{Sin\theta}\, P_{S1}
 &-\frac{R^{2}  G}{2}\, P_{S2}\\
\text{Sym}    &\text{Sym}   & \text{Sym}    &R^{2} G  P_{S3}
\end{pmatrix},
\end{eqnarray}
where $A,\, B,\, C,\, D,\, E,\,F,\,G,\, H,\, J$ and $K$ are the functions of $t$ and $R$, $S$ is the function of $\theta$ and $ \varphi$, $E_{F}^{\pm}=[E \pm F (\partial_{\theta}\hsp^{2}+\frac{1}{2}l(l+1))]S $,  $P_{S1}=(\partial_{\theta}\partial_{\varphi}-cot\theta\partial_{\varphi})S$, $P_{S2}=(\frac{1}{sin\theta}\partial_{\varphi}^{2}+cos\theta \partial_{\theta}
-sin\theta \partial _{\theta}^{2}) S$ and $P_{S3}=(sin\theta\partial_{\theta}\partial_{\varphi}-cos\theta\partial_{\varphi})S$.
Then all perturbation quantities are also divided into  the odd and even parities: 
$l_\mu^B=l_\mu^{BE}+l_\mu^{BO}$, $n_\mu^B=n_\mu^{BE}+n_\mu^{BO}$ and $m_\mu^B=m_\mu^{BE}+ m_\mu^{BO}$, $\Psi_4^{B}=\Psi_4^{BE}+ \Psi_4^{BO}$,  $\lambda^B=\lambda^{BE}+\lambda^{BO}$, $\nu^B=\nu^{BE}+\nu^{BO}$, 
and so on.  Accordingly, we also rewrite the perturbative energy-momentum tensor as 
$T^B_{\mu\nu}=T^{BE}_{\mu\nu}+T^{BO}_{\mu\nu}$. Then, we have $T^B_{n\bar{m}}=T^{BE}_{n\bar{m}}+T^{BO}_{n\bar{m}}$,  $T^B_{\bar{m}\bar{m}}=T^{BE}_{\bar{m}\bar{m}}+T^{BO}_{\bar{m}\bar{m}}$, $T^B_{n n}=T^{BE}_{n n}+T^{BO}_{n n}$.

Taking the Regge-Wheeler gauge
$
  B=F=H=G=0
$ \cite{Thompson},  the null tetrads for the even parity are described by
\begin{align}\label{1.14}
&l_{E}^{\mu}=\Big(\frac{A_0-\varepsilon\, A\,S}{A_0^2},1-\frac{ \varepsilon\,  A_D^+ \,S}{2\, A_0} , \  0, \   0\Big),\notag\\
&n_{E}^{\mu}=\Big(\frac{1}{2},\ \ -\frac{1}{2}\Big(A_{0} +\frac{\varepsilon}{2} \, A_D^- \, S\Big),\ \ 0, \ \ 0 \Big), \notag\\
&m_{E}^{\mu}=\Big(0,\ \   0,\ \  \frac{2+\varepsilon E\, S}{2\,\sqrt{2}\, R},\ \ i \frac{ 2+\varepsilon E\, S}{2\,\sqrt{2}\, R\, sin\theta}\Big), 
\end{align}
where $A_D^{\pm} = A\,\pm A_{0}^2\,  K\mp 2 A_{0}\,  D$. 
The calculation  shows that  
\begin{align}\label{SL}
  \sigma^{BE}=0\;,\qquad\lambda^{BE}=0\;. 
\end{align}
And the null tetrads for the odd parity are given by 
\begin{align}\label{1.15}
&l_{O}^{\mu}=\Big(\frac{1}{A_{0}},\ \ 1, \ \  \frac{\varepsilon\,  C_{J}^{-}}{R \, A_{0}\, sin\theta}\frac{\partial S}{\partial{\varphi}},\ \ -\frac{\varepsilon \, C_{J}^{-}}{R \, A_{0} \,sin\theta}\frac{\partial S}{\partial{\theta} }\Big),\notag\\
&n_{O}^{\mu}=\Big(\frac{1}{2},\ \ -\frac{A_{0}}{2},\ \ \frac{\varepsilon\, C_{J}^{+}}{2\,R\, sin\theta}\frac{\partial S}{\partial{\varphi}},\ \ -\frac{\varepsilon \, C_{J}^{+}}{2\, R\, sin\theta}\frac{\partial S}{\partial{\theta} }\Big),\notag\\
&m_{O}^{\mu}=\Big(0,\ \ 0, \ \ \frac{1}{\sqrt{2}\, R}, \ \ \frac{i }{\sqrt{2}\, R \, sin\theta}\Big),
\end{align}
where $C_{J}^{\pm}=C\,\pm A_{0}J\,$.  For this case, we have   
\begin{align}\label{KN}
  \kappa^{BO}=0\;,\ \ \ \nu^{BO}=0\;.
\end{align}

By substituting  Eq. (\ref{SL}) into Eq. (\ref{Psi4})  and noting that $\bar{\delta}\phi_{11}=0$, we obtain the decoupled equation for the  even parity 
\begin{align}
&\Big[(\Delta+3\gamma-\bar{\gamma}+4\mu+\bar{\mu})(D-\rho) -(\bar{\delta}+3\alpha+\bar{\beta}) (\delta+4\beta)\nonumber \\
&-3\psi_{2}+2\phi_{11}\Big]\psi^{BE}_{4}  =\frac{\hat{\kappa}}{2} \,T_{4}^{BE},  \label{Psi4A}
\end{align}
with 
\begin{align}
T_4^{BE}&=(\Delta+3\gamma-\bar{\gamma}+4\mu+\bar{\mu})[(\bar{\delta}+2\alpha)T^{BE}_{n\bar{m}}  -(\Delta+2\gamma\nonumber \\
&-2\bar{\gamma}+\bar{\mu})T^{BE}_{\bar{m}\bar{m}}]+(\bar{\delta}+3\alpha+\bar{\beta})[(\Delta+2\gamma+2\bar{\mu})T^{BE}_{n\bar{m}}\nonumber \\ &-(\bar{\delta}+2\bar{\beta}+2\alpha)T^{BE}_{nn}],
\end{align}
and from  Eqs. (\ref{KN}) and (\ref{Psi4}) we obtain the equation for the odd parity
\begin{align}
&\Big[(\Delta+3\gamma-\bar{\gamma}+4\mu+\bar{\mu})(D-\rho) -(\bar{\delta}+3\alpha+\bar{\beta}) (\delta+4\beta)\nonumber \\
&-3\psi_{2}-2\phi_{11}\Big]\psi^{BO}_{4}  =\frac{\hat{\kappa}}{2} \, T_{4}^{BO},  \label{Psi4B}
\end{align}
with 
\begin{align}
T_4^{BO}&=(\Delta+3\gamma-\bar{\gamma}+4\mu+\bar{\mu})[(\bar{\delta}+2\alpha)T^{BO}_{n\bar{m}}  -(\Delta+2\gamma\nonumber\\ &-2\bar{\gamma}+\bar{\mu})T^{BO}_{\bar{m}\bar{m}}]+(\bar{\delta}+3\alpha+\bar{\beta})[(\Delta+2\gamma+2\bar{\mu})T^{BO}_{n\bar{m}}\nonumber \\ &-(\bar{\delta}+2\bar{\beta}+2\alpha)T^{BO}_{nn}]-\frac{8}{\hat{\kappa}}\lambda^{BO}(\Delta+\mu+\bar{\mu})\phi_{11}.
\end{align}
We note that Eq. (\ref{Psi4B}) still includes $\lambda^{BO}$. Fortunately, by substituting Eq. ({\ref{KN}}) into Eq.(3.8.11j) in Ref. \cite{Carmeli},  we obtain 
\begin{align}\label{LPsi}
  (\Delta+\mu+\bar{\mu}+3\gamma-\bar{\gamma})\lambda^{BO}+\psi_{4}^{BO}=0\;.
\end{align}
Using Eqs. (\ref{Psi4B}) and (\ref{LPsi}) we can find out the decoupled equation for the  odd parity $\psi_{4}^{BO}$. 

\vspace{0.3cm}

\subsection{The rate of energy-loss, radiation-reaction force and waveform for the ``plus" and ``cross" modes}

In general, the solution of $\psi_4^B$ can be expressed as $
\psi_4^B= Re(\psi^{BE}_4+\psi^{BO}_4)-  i Im(\psi^{BE}_4+\psi^{BO}_4)$,  which implies that
$
\ddot{h}_{+}=Re(\psi^{BE}_4+\psi^{BO}_4) $ and $
\ddot{h}_{\times}=Im(\psi^{BE}_4+\psi^{BO}_4).
$
Therefore, the energy-loss rate can be expressed as  \cite{Magg}
\begin{eqnarray}
\frac{d E[g_{\mu\nu}^{\text{eff}}]}{dt }&=&\int \frac{c^3}{16\pi G}(\dot{h}_{+}^2+\dot{h}_{\times}^2)r^2 d \Omega\nonumber \\
&=&\frac{c^3}{16\pi G\omega^2}\int \Big\{\Big[Re(\psi^{BE}_4+\psi^{BO}_4)\Big]^2\nonumber \\
&+& \Big[Im(\psi^{BE}_4 +\psi^{BO}_4)\Big]^2\Big\}r^2 d \Omega.\label{ERate}
\end{eqnarray}

With the energy-loss rate at hand,  we can then simply obtain the radiation-reaction force from Eq. (\ref{dEt}).

It is well known that the ``plus" and ``cross" modes of the gravitational wave can be expressed in terms of spin-weighted $s=-2$ spherical harmonics \cite{Kidder}, i.e., 
\begin{align}
h_{+}-i h_{\times}=\sum_{l=2}^{\infty} \sum_{m=-l}^{l} h^{lm} \  _{-2}Y^{lm}(\theta, \varphi) .
\end{align} 
Thus, by comparing it with $\psi^B_{4}=\ddot{h}_{+}-i\ddot{h}_{\times}$,    
we can read out the waveform. And then, following Refs.  \cite{Pan2,Pan3,Damour2009}, we could rewrite the inspiral-plunge modes. 

For a complete self-consistent EOB theory, we should carry out the concrete calculations for the radiation-reaction force by using the method proposed in Refs. 
\cite{Ref:poisson,CutlerPRD,Sasaki17,TagoshiSasaki745}, and  then rewrite the inspiral-plunge modes following Refs.  \cite{Pan2,Pan3,Damour2009}. However, these  calculations are very tedious which we will present elsewhere.

\section{Conclusions}
\vspace{0.3cm}

To summary, form the Hamilton equations (\ref{HEq}) for the EOB system we know that  the  Hamiltonian $H$, the radiation-reaction force  $ {\cal F}_\varphi$  and the waveform for the ``plus" and ``cross" modes should be based on the same EOB model, i.e.,  a test particle with rest mass $m_0$ moving in an effective background spacetime $g_{\mu\nu}^{\rm eff}$ with the mass $M_0$.
To build such a  self-consistent EOB theory, the key step is to look for the decoupled equation for the  component of perturbed Weyl tensor $\psi^B_{4}=\ddot{h}_{+}-i\ddot{h}_{\times}$ in the effective background spacetime $g_{\mu\nu}^{\rm eff}$ obtained by means of the scattering angles based on the PM approximation for the non-spinning  binaries. This sounds like a difficult task, since there are non-vanishing tetrad components of the tracefree Ricci tensor  ($\phi_{00},$ $ \phi_{22} $ and $\phi_{11}$) for the effective spacetime (\ref{Mmetric}). By using the freedom for the parameters ($a_i, \ d_i$) in the effective metric,  i..e., there are two unknowns $a_i$ and $d_i$ for one  equation at each $i$th  PM approximation order, we can take $ D_0=1$ which indicates  that  $\phi_{00}=0$ and $ \phi_{22}=0$. Then, by dividing the perturbation part of the metric into  the odd and even parities, we fortunately find  the decoupled equations for the  even parity  $\psi^{BE}_{4}$ and the odd parity  $\psi^{BO}_{4}$, respectively. 

With the decoupled equations for $\psi^B_{4}$  at hand, we can find the radiation-reaction force which can be derived from the  rate of energy-loss for the ``plus" and ``cross" modes by using the method proposed by Poisson, Sasaki, and Tagoshi et. al  in Refs. 
\cite{Ref:poisson,CutlerPRD,Sasaki17,TagoshiSasaki745}, and we can also read out the waveform  for these modes and then rewrite the inspiral-plunge modes following Pan and Damour et. al  \cite{Pan2,Pan3,Damour2009}. 
Therefore, by substituting the radiation-reaction force and Hamiltonian (\ref{Heob}) into the Hamilton equations (\ref{HEq}), we set up  a frame for the self-consistent EOB theory in which the Hamiltonian $H[g_{\mu\nu}^{\text{eff}}]$, the radiation-reaction force  $ {\cal F}_\varphi [g_{\mu\nu}^{\text{eff}}]$  and the waveform for the ``plus" and ``cross" modes of the gravitational wave are based on the same effective background metric. 
In the next step we would like to carry out the concrete calculations and construct a  gravitational waveform template based on the self-consistent EOB theory  for the spinless  binaries. Further more, we will try to set up a self-consistent EOB theory for spin  binaries   based on  post-Minkowskian approximation. 

\vspace{0.3cm}
\acknowledgments
{\it Acknowledgments}: We would like to thank professors S. Chen and Q. Pan  for useful discussions on the manuscript. This work was supported by the Grant of NSFC Nos. 12035005, 12122504 and 11875025, and National Key Research and Development  Program of China No. 2020YFC2201400.  

\bibliography{refs}

\end{document}